\documentclass[aps,pra,twocolumn,groupedaddress,showpacs]{revtex4}
\usepackage{epsfig}

\def\q{{\bf q}}
\def\p{{\bf p}}
\def\k{{\bf k}}

\def\8{\infty}
\def\oh{\frac{1}{2}}

\def\undertext#1{\vtop{\hbox{#1}\kern 1pt \hrule}}

\def\be{\begin{equation}}
\def\ee{\end{equation}}
\def\bea{\begin{eqnarray} & &}
\def\eea{\end{eqnarray}}

\def\rf#1{(\ref{#1})}

\def\rf#1{(\ref{#1})}

\def\rfs#1{Eq.~\rf{#1}}

\def\nn{\nonumber}

\begin{document}


\title{Stability of the fermionic gases close to a $p$-wave Feshbach resonance}


\author{J. Levinsen$^{1}$, N. R. Cooper$^{2}$ and V. Gurarie$^{3}$}
\affiliation{$^1$Laboratoire Physique Th\'eorique et Mod\`eles Statistique, Universit\'e Paris Sud, CNRS, 91405 Orsay, France\\
$^2$T.C.M. Group, Cavendish Laboratory, J. J. Thomson Avenue, Cambridge CB3 0HE, United Kingdom \\
$^3$Department of Physics, University of Colorado,
Boulder CO 80309, USA}


\date{\today}

\begin{abstract}
We study the stability of the paired fermionic $p$-wave superfluid made out of identical atoms all in the same hyperfine state close to 
a $p$-wave Feshbach resonance. First we reproduce known results concerning the lifetime of a 3D superfluid, in particular, we show
that it decays at the same rate as its interaction energy, thus precluding its equilibration before it decays. Then
we proceed to study its stability in case when the superfluid is confined to 2D by means of an optical harmonic potential. We find
that the relative stability is somewhat improved in 2D in the BCS regime, such that the decay rate is now slower than the appropriate interaction
energy scale. The improvement in stability, however, is not dramatic and one probably needs to look for other mechanisms
to suppress decay to create a long lived 2D $p$-wave fermionic superfluid. 
\end{abstract}

\maketitle

\section{Introduction}

Recent success with the BEC-BCS crossover experiments in the atomic fermionic gases with $s$-wave Feshbach resonances \cite{Jin2004,Ketterle2004,Hulet2005} inspired studies towards creating $p$-wave fermionic superfluids using $p$-wave Feshbach resonances~\cite{Botelho2005,Ho2005,Gurarie2005,Yip2005}. A number of novel features made the $p$-wave superfluids attractive, as discussed at length in Ref.~\cite{Gurarie2007}. First of all, it is enough to put atoms into identical hyperfine states to suppress their $s$-wave scattering, leaving $p$-wave scattering as the strongest scattering channel. And indeed, $p$-wave Feshbach resonances between atoms in identical hyperfine states had been identified and studied some time  ago \cite{Regal2003,Ticknor2004}. Next, the $p$-wave superfluids have a number of features distinguishing them from their $s$-wave counterparts. A richer $p$-wave order parameter allows for a possibility of observing different phases of the $p$-wave superfluids, some of which are akin to the phases of superfluid Helium III \cite{Wolfe}. Chiral and polar phases of the $p$-wave condensates are possible, which differ by the projection of the angular momentum of the Cooper pairs (or molecules) of the condensate onto the chosen axis. If that projection is $\pm 1$, the condensate is called chiral, and if it is $0$, the condensate is called polar. Another important feature is that as the system is tuned from BCS to BEC, it does not go through a crossover as in the $s$-wave case, but rather goes through a phase transition, as was first discussed by G. Volovik long before current experiments on the BCS-BEC systems became possible, Ref.~\cite{VolovikBook}. Thus the BCS and BEC are two distinct phases of $p$-wave condensates (with either of them possibly being chiral or polar, bringing the total number of phases to four). Finally, when confined to 2D, the chiral BCS phase of the $p$-wave superfluids is topological and its vortices have trapped quasiparticles which obey non-Abelian statistics \cite{Read2000,Ivanov2001}. Such quasiparticles were suggested to be used as  topologically protected qubits  to construct decoherence free quantum computers \cite{Kitaev2003}. 

However, the program to create these superfluids suffered a setback when experimental studies of the $p$-wave Feshbach molecules showed they were unstable, with the lifetime varying between 2 and 20 $ms$ \cite{Salomon2004,Gaebler2007,Ticknor2008,Inada2008}. Although some of these studies were done with molecules made of atoms of $^{40}$K, which are inherently unstable due to dipolar relaxation \cite{Gaebler2007}, the rest of the studies used atoms of $^6$Li, whose $p$-wave molecules should not, by themselves, exhibit any instability. 

A common mechanism which can lead to instability in atomic gases is the atom-molecule and molecule-molecule relaxation \cite{Petrov2005}. This is the process which, for example, involves one of the atoms approaching a molecule, with the result being that the molecule collapses into one of its strongly bound states, while the excess energy is carried away by the atom. Such processes are suppressed in the $s$-wave superfluid due to the Pauli principle, as was convincingly demonstrated in Refs.~\cite{Petrov2004,Petrov2005}. However, the Pauli principle does not protect the $p$-wave superfluids, potentially leading to much shorter lifetimes. 

Refs.~\cite{Levinsen2007,Lasinio2008}  examined the stability of the $p$-wave condensates due to these relaxation processes, as well as due to a possible 
recombination into trimers \cite{Levinsen2007,Levinsen2007a,Lasinio2008}. We established that the decay rate of a condensate of $p$-wave molecules close to Feshbach resonance is given by
\be \label{eq:1}
\Gamma_{3D}  \sim \frac{\hbar}{m \ell^2} \frac{R_e}{\ell},
\ee
where $\ell$ is the typical interatomic spacing and $R_e$ is the van der Vaals length (the interaction range), typically estimated to be $\sim50$ au (so the ratio $\ell/R_e$, assuming that $\ell \sim 10000$ au, is of the order of $200$). $\hbar^2/(m \ell^2)$ is the Fermi energy of the gas. Thus for a gas of Fermi energy about 10KHz, this gives an estimate of the lifetime of $1/\Gamma \sim 20$ms, which is not far from what is measured experimentally. This should be compared with the corresponding expression for the decay rate of the $s$-wave condensate, \be \label{eq:2} \Gamma_{s-wave} \sim \frac{\hbar}{m\ell^2} \left( \frac{R_e}{\ell} \right)^{3.55},
\ee
which is orders of magnitude slower than the $p$-wave rate, leading to a stable condensate. 

The calculations leading to the expression \rfs{eq:1} were done solely in three dimensional space. Yet the most interesting $p$-wave condensate, the one with non-Abelian quasiparticles, has to be confined to two dimensions. The confinement may affect the lifetime of the condensate. In the absence of any experiments where the $p$-wave resonant gases are confined to 2D, it is imperative that a theoretical calculation is done estimating this lifetime. 

In this paper we estimate the lifetime of the $p$-wave condensates close to Feshbach resonance confined to 2D. For purely 2D condensates we find that their decay rate is given by 
\be \label{eq:3} 
\Gamma_{2D} \sim \frac{\hbar}{m\ell^2}.
\ee
This is even faster than the 3D case, \rfs{eq:1}. However, for the quasi-2D condensates, the ones which are confined to a ``pancake'' of width $d$, where $d \ll \ell$ and at the same time $d \gg R_e$, we find that the decay rate is given by
\be \label{eq:4}
\Gamma_{quasi-2D} \sim \frac{\hbar}{m\ell^2} \frac{R_e}{d}.
\ee
Notice that this expression interpolates between Eq.~\rf{eq:1} and \rf{eq:3}. Indeed, as $d$ becomes smaller than $\ell$, \rfs{eq:1} gets replaced by \rfs{eq:4}. As $d$ is decreased, it eventually becomes smaller than $R_e$, at which point \rfs{eq:4} gets replaced by \rfs{eq:3}. 

The rate in the quasi-2D geometry, given by \rfs{eq:4} is somewhat faster than the 3D rate \rfs{eq:1}. So one may jump to the conclusion that the quasi-2D geometry in fact decreases the lifetime of the condensate. This however must be contrasted with the fact that in 2D the interactions are stronger. And indeed, the typical interaction energy per particle of the 3D condensate (assuming that it is in the ``strong'' resonance regime \cite{Levinsen2007}
and concentrating, for simplicity, on the BCS regime only)
\be
E_{3D} \sim \frac{\hbar^2}{m\ell^2} \frac{R_e}{\ell},
\ee
that is it is of the same order as $\Gamma_{3D}$ from \rfs{eq:1}. Thus the condensate decays in 3D as fast as it interacts.  On the other hand, in $2D$ the interaction energy is
given by
\be  \label{eq:6} E_{2D} \sim \frac{\hbar^2}{m\ell^2} \frac{1}{\log \left[ \frac{\ell}{R_e} \right] }.
\ee
Comparing it with $\Gamma_{2D}$ from \rfs{eq:3}, we see that the interaction energy is weaker than the decay rate, by a logarithmic factor. 

However, the case we are interested in is quasi-2D, when the condensate is confined to a pancake of width $d$. Under these conditions, the decay rate is
given by \rfs{eq:4}, while the interaction strength is still given by  \rfs{eq:6}, with $d$ substituted for $R_e$. Provided that 
\be \label{eq:7}
\frac{1}{\log \left[ \frac{\ell}{d} \right] } \gg \frac{R_e}{d},
\ee
the interaction energy in quasi-2D can be larger than the decay rate, thus creating a situation where the condensate might have sufficient time to form. 
In turn, since $d \gg R_e$ and logarithms, even of large arguments, are typically not very large, \rfs{eq:7} may indeed hold. The fact that the quasi-2D BCS $p$-wave condensates are more stable than their 3D counterparts is the main conclusion of this paper. 

The rest of the paper is organized as follows. 

In section \ref{sec:II} we go over the analysis of the stability of the 3D $p$-wave superfluid, mostly following discussions in Ref.~\cite{Levinsen2007}. In particular, in subsection \ref{sec:IIA} we study the two channel model describing the superfluid and explain the difference between strong and weak resonances, as well as narrow and wide ones, while in subsection \ref{sec:IIB} we go over the stability analysis. 

In section III we present the analysis of the stability in 2D. First, in subsection \ref{sec:IIIA} we set up a 2D $p$-wave gas. Next, in subsection \ref{sec:IIIB} we discuss the stability of the condensates when the molecules are large, relevant in 3D $s$-wave and 2D $p$-wave cases. In the next subsection
\ref{sec:IIIC} we set up the three-body problem which needs to be solved to compute the decay rate. In subsection \ref{sec:IIID} this problem
is solved. Finally, in subsection \ref{sec:IIIE} the implications of the solution are discussed and the decay rate is derived.  Subsections
\ref{sec:IIIA} and \ref{sec:IIID} are the most technically involved parts of the paper and can be safely omitted at first reading. 

In section IV, we go over the analysis in quasi-2D, where the condensate is confined to a pancake geometry. This section is followed by Conclusions
and two Appendices. 

Our final remark concerns the usage of the Planck constant $\hbar$. It generally helps to omit it in calculations because it clutters the expressions and
makes them harder to manipulate, while it can always be restored everywhere by dimensional analysis. So we adopt notations where  $\hbar=1$ everywhere in this paper from here on.




\section{Stability of 3D $p$-wave fermionic superfluid}
\label{sec:II}
\subsection{The two channel model}
\label{sec:IIA}
To describe the 3D $p$-wave resonantly coupled superfluid we employ a
two-channel model with Hamiltonian
\cite{Timmermans1999,Timmermans2001,Holland2001,Gurarie2005,Yip2005,
  Gurarie2007}
\begin{eqnarray}
 H &=& \sum_\p \frac{p^2}{2m} ~\hat a^\dagger_\p
\hat a_{\bf p} + \sum_{{\bf q}, \mu} \left(\epsilon_0 +
{\frac{q^2}{4m}} \right)
\hat b_{\mu{\bf q}}^\dagger \hat b_{\mu {\bf q}}\cr
&+&\sum_{{\bf p},{\bf q},\mu} ~{g ( \left| \p \right|) \over \sqrt{V}} \left( \hat b_{\mu
{\bf q}} ~p_\mu~ \hat a^\dagger_{{\q\over 2}+\p} ~\hat a^\dagger_{{\q
\over 2}-\p} + h. c. \right).
\label{eq:H}
\end{eqnarray}
Here $\hat a^\dagger$ and $\hat a$ are creation and annihilation
operators of a (spinless) fermion with mass $m$, while the bare spin 1
bosonic diatomic molecule is created and annihilated by $\hat
b^\dagger_\mu$ and $\hat b_\mu$. The vector index
$\mu$ represents the projection of spin on some axis.

The superfluid described by the Hamiltonian (\ref{eq:H}) depends on
four parameters. Of these, $\epsilon_0$ is the bare detuning,
controlling the position of the Feshbach resonance, while the particle
number $N$  is the expectation value of the operator
\begin{equation}
\hat N = \sum_\p \hat a^\dagger_\p \hat a_\p + 2
\sum_{\mu, {\bf q}} \hat b^\dagger_{\mu \q} \hat b_{\mu \q}.
\end{equation}
Often it is convenient to trade the particle number for the Fermi energy $\epsilon_F$. The Fermi energy
is defined as the energy of a free Fermi gas whose particle number concides with $N$ above,
and is explicitly given by
\be
\epsilon_F = \frac{(6 \pi^2 N)^{\frac 2 3}}{2 m V^{\frac 2 3}}.
\ee
Another way to represent the particle number is by the interparticle spacing 
\be
\ell = \left( \frac{V}{N} \right)^{\frac 1 3} = \frac{1}{n^{\frac 1 3}}
\ee
where $n$ denotes the density of particles $n=N/V$. 

The interaction term of the Hamiltonian turns two fermions into a
boson and vice versa. It is described by the interaction strength
$g(|\p|)$, the momentum dependence of which originates in the fact
that the interaction is not point-like, rather the interaction
strength is proportional to the momentum space wavefunction of the
molecule. As such, the interaction needs to be supplemented with a
cut-off of order $\Lambda\sim 1/R_e$ where $R_e$ is the physical
size of the molecule, or the interaction range as discussed in the Introduction. We will write the interaction strength as
\begin{equation}
g(|\p|)=g\,\xi(|\p|/\Lambda).
\end{equation}
where $g$ is the asymptotic value of the interaction strength for
small momenta and the dimensionless function $\xi$ describes the
quick fall-off of the interaction strength for momenta
$|\p|\gg\Lambda$. In this paper, for simplicity we will let
\begin{equation}
\xi(x)=\Theta(1-x)=\left\{\begin{array}{cc}
1, & x<1  \\ 0, & x>1\end{array}\right. .
\end{equation}
with $\Theta$ the usual stepfunction. The precise shape of the cut-off
function does not affect the conclusions of this paper, see the
discussion in Section \ref{sec:2dscat}. The parameters $\epsilon_0, \epsilon_F,g,\Lambda$ completely 
characterize the $p$-wave superfluid. 

Two important dimensionless combinations can be constructed out of these parameters. One is given by
\be \gamma \sim \frac{m^2 g^2}{\ell}.
\ee
If this parameter is small, $\gamma \ll 1$, then mean field theory can safely be employed to analyze the Hamiltonian \rfs{eq:H}. In a typical experiment
$\gamma$ is indeed small, being on the order of $\gamma \sim 1/10$ \cite{Gurarie2005,Gurarie2007}. By analogy with $s$-wave Feshbach resonances, the  case of small $\gamma$ can be termed that of narrow resonance (while the experimentally irrelevant case of $\gamma \gg 1$ can be termed broad $p$-wave 
resonance).

The second parameter,
\be \label{eq:15} c_2 = \frac{m^2 g^2 \Lambda}{3\pi^2} \ee
can also be formed.  Following Ref.~\cite{Levinsen2007} we term the superfluid with large $c_2$ the case of strong $p$-wave resonance, and correspondingly the case with small $c_2$ weak resonance. In current experiments, $c_2$ is typically large, thus the resonances studied so far were strong 
\cite{Gurarie2007}.

 It should be emphasized that under the condition $c_2 \gg 1$, it is possible to trade \cite{Gurarie2007} the two-channel model \rf{eq:H} for the one-channel model
  \begin{eqnarray}
\label{eq:1c}  &H_{1-c} = \sum_\p \frac{p^2}{2m} ~\hat a^\dagger_\p
\hat a_{\bf p} - & \\ & 
 \sum_{{\bf p},{\bf p'},{\bf q},\mu} ~\frac{ g ( \left| \p \right|) g ( \left| \p' \right|) } {V \epsilon_0} p_\mu p'_\mu ~ \hat a^\dagger_{{\q\over 2}+\p} ~\hat a^\dagger_{{\q
\over 2}-\p} \hat a_{\frac {\bf q} 2 - \p'} \hat a_{\frac \q 2 + \p' }.& \nonumber
\end{eqnarray}
While $c_2$ is indeed large in current experiments, we prefer to employ for our analysis the two-channel \rfs{eq:H}. Indeed, to analyze \rf{eq:1c}, one typically needs
to employ a Hubbard-Stratonovich transformation to turn it into something similar to \rfs{eq:H} first, and proceed from there. We find it more straightforward to work directly with \rfs{eq:H}. 

To further elucidate the meaning of the model \rfs{eq:H} and its relation to real phenomena, we consider the scattering of two atoms with
momenta ${\bf k}$ and $-{\bf k}$ into momenta ${\bf k}'$ and $-{\bf k}'$. This scattering proceeds via the $p$-wave channel and its scattering amplitude is given by \cite{LL} (we emphasize that
this is a partial scattering amplitude, while the full amplitude is given by the standard expression $3 f_1(k) P_1(\cos \theta)$ where $\theta$ is the angle between
the incoming and outgoing momenta)
\be \label{eq:17}
f_1(k) = \frac{k^2}{-\frac{1}{v} + \frac{k_0}{2} k^2 - i k^3}.
\ee
Here $v$ is called the scattering volume. In terms of the parameters of \rfs{eq:H} it is given by \cite{Gurarie2007}
\be \label{eq:18} v^{-1} = - \frac{6 \pi \omega_0 (1+c_2)}{mg^2}, \ \omega_0 = \frac{\epsilon_0- \frac{m \Lambda^3 g^2}{9 \pi^2}}
{1+c_2}.
\ee
In an experiment, $\epsilon_0$ (and thus $\omega_0$) is tuned by varying the magnetic field. This induces a change in $v$, with $1/v$ crossing zero as $\epsilon_0$ is decreased (in this regard, the behavior of the
scattering volume $v$ is completely equivalent to the behavior of the scattering length $a$ in an
$s$-wave Feshbach resonant scattering). 

The parameter $k_0$ replaces the ``effective range'' parameter $r_0$ of the $s$-wave resonances and is given by
\be
k_0 = - \frac{12 \pi}{m^2 g^2} \left(1 + c_2 \right).
\ee
The poles of the scattering amplitudes describe resonant scattering at positive $\omega_0$ and bound states (molecules) at
negative $\omega_0$. These occur at (neglecting the $ik^3$ term in the denominator of \rfs{eq:17}, small at small $\omega_0$)
\be \label{eq:21}
\frac{k^2}{m}  \approx \omega_0.
\ee
This elucidates the meaning of $\omega_0$ introduced in \rfs{eq:18}. One can also remark that the total scattering cross section, for the $p$-wave
scattering, is given by
\cite{LL}
\be
\sigma = 12 \pi \left| f_1(k) \right|^2.
\ee

$\omega_0$, as long as it is positive, can be measured in an experiment by looking at the energy of colliding particles at which the scattering cross section 
$\sigma$ has a maximum. At the same time, $\epsilon_0$ is varied by the magnetic field according to
\be
\omega_0 \sim \frac{ \mu_B \left(B-B_0 \right)}{1+c_2},
\ee
where $B_0$ is the magnetic field corresponding to the resonance and $\mu_B$ is the effective Bohr magneton. Thus measuring $\omega_0$ and
$B-B_0$ simultaneously allows to determine whether the resonance is weak or strong ($c_2 \ll 1$ or $c_2 \gg 1$). Strong resonances will appear as 
the ones where the slope of the curve $\omega_0$ vs $\mu_B \left(B-B_0 \right)$ is small. 

The dependence of the scattering volume $v$ on the magnetic field $B$ is given by
\be
v = -\frac{m g^2}{6 \pi \mu_B \left(B-B_0 \right)},
\ee
reminiscent of the magnetic field dependence of the scattering length in an $s$-wave Feshbach resonance experiment. 
We notice that if one tunes the magnetic field $B$ off resonance by the amount such that $\mu \left(B - B_0 \right)$ 
is equal to the Fermi energy of the gas, then the scattering volume will be much smaller than the cube of the interparticle
spacing $\ell^3$ if the resonance is narrow ($\gamma \ll 1$), and much larger than the cube of the spacing if the
resonance is broad ($\gamma \gg 1$). This justifies the name ``narrow'' vs ``broad''. Indeed, in case of the
broad resonance, $v$ vs $B-B_0$ graph will appear much broader than in the case of narrow resonance. 
Notice the complete equivalence of this term with its $s$-wave counterpart
\cite{bruun:140404,Cornell2004a,Ho2004,Andreev2004,Petrov2004a,Gurarie2005,Gurarie2007}.

We emphasize that other  authors, 
such as the ones of Ref.~\cite{Lasinio2008}, prefer to restrict  the usage of the term ``broad'' vs ``narrow'' for $c_2$ being large or small.
 While the concrete terminology is a matter of taste,  $\gamma$ is a parameter
 which  more accurately reflects the notion of ``broad'' vs ``narrow'', as these terms are used in the $s$-wave resonance context. 

\subsection{The analysis of the stability of the 3D $p$-wave condensate}
\label{sec:IIB}

Here we reproduce the analysis of the stability of the 3D $p$-wave condensate from Ref.~\cite{Levinsen2007}. Suppose, at some negative $\omega_0$, 
$p$-wave molecules form whose binding energy is $\omega_0$. The radial part of their wavefunction, at distances much larger than the interaction range $R_e \sim 1/\Lambda$, is given by \cite{LL,Ticknor2008}
\be \label{eq:24}
\Psi(r) = \frac{e^{-\kappa r}}r \left( 1+ \frac{1}{\kappa r} \right),
\ee
where $\kappa=\sqrt{m \left| \omega_0 \right|}$. At distances much smaller than $1/\kappa$ but still much larger than $R_e$, $R_e \ll r \ll 1/\kappa$, the wavefunction can be well approximated by
\be \label{eq:25} \Psi(r) \sim \frac{1}{r^2}.
\ee
We notice that the normalization condition of the wavefunction
\be
\int_{R_e}^{\frac 1 \kappa} r^2 dr \left| \Psi(r) \right|^2 \sim \frac{1}{R_e}
\ee
leads to the following normalized expression
\be
\Psi(r) = \frac{1}{ r^2 \sqrt{R_e}}
\ee
independent of $\omega_0$. 
In other words, most of the weight of the wavefunction is concentrated at distances $R_e$, or the molecules are small. 

Suppose two such molecules collide. It is possible that the collision will lead to one molecule forming a strongly bound state, while the atoms of the
other molecules absorb the energy and fly apart. It is also possible that three of the atoms form a strongly bound trimer, while the remaining atom flies away \cite{Lasinio2008,Levinsen2007}. The rate of this process can be estimated as follows. The total rate is given by
\be \label{eq:28}
\Gamma \sim n \sigma v,
\ee
where $n=1/\ell^3$ is the density of particles, $v$ is their velocity, and $\sigma$ is the inelastic cross section for this process. In turn, $\sigma$ can be estimated as  a product of the elastic cross section  of two small objects size $R_e$ each, or $R_e^2$ times the time they spend in the vicinity $R_e/v$ times the rate of the decay. That rate may be hard to calculate, but it must be of the order $1/(m R_e^2)$ by dimensional analysis. Putting this all together we find
\be  \label{eq:29}
\Gamma \sim \frac{1}{\ell^3} R_e^2 \frac{R_e}{v} \frac{1}{mR_e^2} v = \frac{1}{m \ell^2} \frac{R_e}{\ell},
\ee
which matches \rfs{eq:1} given in the Introduction. 

Another simple way to derive $\Gamma$ proceeds as follows. Due to the ultraviolet divergencies in the $p$-wave two-channel model \rfs{eq:H}, any 
transition amplitude, elastic or inelastic, between some states of two molecules or a molecule and an atom, should go as $R_e$ (up to a phase). The inelastic cross
section can then be estimated as \cite{LL}
\be
\sigma_{in} \sim R_e^2 \frac{k_f}{k_i},
\ee
where $k_f=1/R_e$ is the final momentum of the particle, and $k_i=mv$ where $v$ is the incident velocity. This leads to
\be
\Gamma \sim \sigma_{in} n v \sim \frac{1}{m\ell^2} \frac{R_e}{\ell},
\ee which coincides with \rfs{eq:29}. This derivation makes it clear that \rfs{eq:29} applies not only to molecule-molecule, but also molecule-atom collisions. 

On the BCS side of the resonance, one could argue that the atoms spend some fraction of their time virtually forming molecules. Those will
also undergo atom-molecule relaxation so the decay rate \rfs{eq:29} applies here as well. One can also consider direct 3 body recombination of atoms. Experimental and theoretical
studies \cite{Regal2003,Greene2003} lead to a rate numerically close to  the one estimated here for the molecule-molecule and molecule-atom
relaxation, at the relevant densities. 

Is the decay rate given by \rfs{eq:29} too fast, or sufficiently slow?
Obviously it has to be slow enough for the condensate to form and equilibrate before it decays. Issues of equilibration are subtle. Instead of considering them, we estimate the interaction rate between the atoms in the BCS phase  and between the molecules in the BEC phase. This rate corresponds to a time scale which is clearly shorter than the equilibration time. Thus, if this energy scale is larger than the decay rate, there is a chance that the condensate would have time to equilibrate. 

In the BCS regime, the molecules have positive energy and if left in vacuum decay into atoms. The molecular decay rate, computed at the Fermi energy, corresponds to the time scale at which atoms interact. This rate can be read off the scattering amplitude \rfs{eq:17}. For that, we find the pole of this scattering amplitude, not just the real part as in \rfs{eq:21}, but also with its imaginary part. It is given by
\be \label{eq:30}
\frac{k^2}{m} \approx \omega_0 - \frac{i}{6\pi} \frac{m^{\frac 5 2} g^2}{1+c_2} \omega_0^{\frac 3 2}.
\ee
The imaginary part is the decay rate. 

We estimate it in case of strong resonances, or $c_2 \gg 1$. Substituting $c_2$ from its definition \rfs{eq:15} and choosing $\omega_0$ to be the Fermi energy of the gas, or $\omega_0 \sim 1/(m \ell^2)$, we find the interaction energy
\be \label{eq:31}
E_{3D} \sim \frac{1}{m \ell^2} \frac{R_e}{\ell}.
\ee
Note that the ratio $g^2/(1+c_2)$ is an {\sl increasing} function of $c_2$. Thus for weak resonances when $c_2 \ll 1$, the interaction energy would be even weaker than the one given by \rfs{eq:31}. So the case of the strong resonance corresponds to the strongest possible interactions. This, combined with
the fact that the $p$-wave Feshbach resonances experimentally studied so far appear to be strong, prompts us to concentrate in this paper mostly
on strong resonance. 

Yet as we see this interaction has the same functional form as the decay rate \rfs{eq:29}. Of course, these are estimates of these quantities and perhaps numerical coefficients
omitted in Eqs.~\rf{eq:31} and \rf{eq:29} conspire to make one larger than the other. But typically we expect that the decay rate and the interaction energy are of the same order, precluding the formation of the condensate before it decays even in case of strong resonance. 

Another way to estimate the interaction rate is in the BEC regime. With the elastic scattering cross section being $R_e^2$, the elastic scattering rate
is
\be \label{eq:3DBEC} E_{3D, BEC} \sim R_e^2 n v \sim E_{3D} {mv R_e} \sim E_{3D} \frac{R_e}{\ell},
\ee
since $mv$ is the characteristic momentum of the molecules, which is roughly equal to $1/\ell$. Obviously, $R_e/\ell \ll 1$, thus the interaction
rate between the molecules in the BEC regime is even slower than the interaction rate between the atoms in the BCS regime. In particular, it
is slower than the decay rate of the molecules, thus making the observation of the 3D BEC $p$-wave condensate even less likely than its
BCS counterpart. 


We can compare the estimates derived here with the measured decay constants from Ref.~\cite{Inada2008} (see their Table I). The decay constant is defined as 
\be \label{eq:35} K = \Gamma/n,
\ee and for the estimate \rfs{eq:29} gives (for once, we explicitly reintroduce the constant $\hbar$)
\be
K = \frac{\hbar R_e}{m} \approx 3 \cdot 10^{-11} cm^3 s^{-1}.
\ee
Here we take $R_e = 50$ au, and take $m$ to be the mass of $^6$Li. 
This is very close to the measured atom-molecule decay constant and is one order of magnitude smaller than the measured molecule-molecule
decay constant. We do not know why the measured molecule-molecule decay constant is faster by a factor of 10, but note that the derivation presented here
ignores the details of the short range physics and could easily be off by a factor of 10. 

We also note that Ref.~\cite{Inada2008} quotes that the elastic scattering rate between the molecules is faster than the inelastic rate, while the
estimates presented here point towards the elastic rate being slower than the inelastic rate. We do not know the reason for this discrepancy.

The conclusion is, the 3D $p$-wave superfluids decay as fast as they interact and do not have time to form before they decay. 

\section{The $p$-wave superfluid in two dimensions
\label{sec:2dscat} }

\subsection{The two channel model}
\label{sec:IIIA}
We now consider the case of the 2D $p$-wave superfluid, governed by the same two-channel model \rfs{eq:H}, but in two dimensional space. 
Reducing the space dimensionality changes the dimension of the coupling. In fact, the $p$-wave two channel model's upper critical
dimension is $2$ \cite{Sachdev2007,CommentDim}. The coupling 
constant $g$ is now a dimensionless quantity. The linear divergence which led to the appearance of $c_2$ in the 3D calculations is now replaced by
a logarithmic divergence. 

As in 3D, it is instructive to compute the scattering amplitude of two atoms. Since to our knowledge  this was not done before in the literature for the two-channel model, 
here is its  derivation.

\begin{figure}[ht]
\includegraphics[height=.7 in]{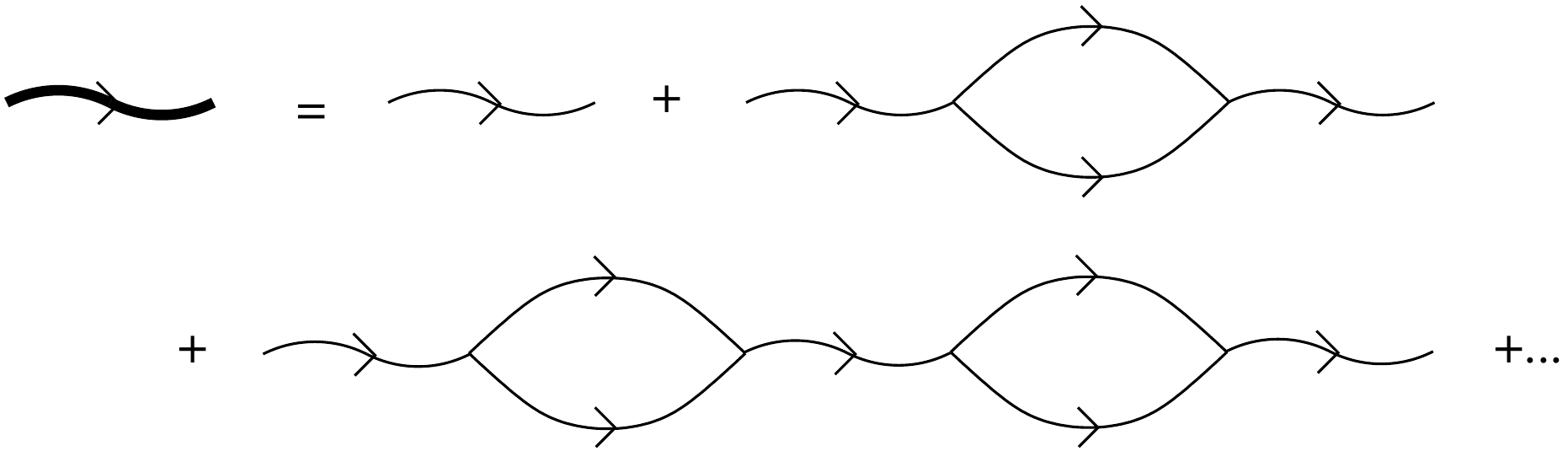}
\caption{The renormalized propagator of spin 1 molecules. Here the standard notations for the two-channel model \rfs{eq:H} are used: the thin straight lines
are fermionic propagators, the thin wavy lines are bare bosonic propagators, and the thick wavy lines are full bosonic propagators.}
\label{fig:bosonprop}
\end{figure}

According to the Hamiltonian (\ref{eq:H}) the propagator of fermionic
atoms is the free propagator
\begin{equation} \label{eq:34}
G(p) = \frac1{p_0-p^2/2m+i0}.
\end{equation}
For simplicity of notation, $p$ is used both as the three-vector
$(\p,p_0)$ and as the absolute value of the momentum $|\p|$. The
bare propagator of the bosonic spin 1 molecules is equal to
\begin{equation}
D^0_{\mu\nu}(p) = \frac{\delta_{\mu\nu}}{p_0-p^2/2m-\epsilon_0+i0}
\equiv D^0(p)\delta_{\mu\nu}.
\end{equation}
The molecular propagator should be renormalized by the
presence of fermionic loops as depicted in
Fig.~\ref{fig:bosonprop}. The fermionic loop separating molecules of
spin $\mu$ and $\nu$ is diagonal in spin indices and is denoted
$\Sigma_{\mu\nu}\equiv\Sigma\delta_{\mu\nu}$. The full propagator is
then given by
\begin{equation}
D_{\mu\nu}(p) = \frac{\delta_{\mu\nu}}{D^0(p)-\Sigma(p)}
\equiv D(p)\delta_{\mu\nu},
\end{equation}
with the fermionic loop taking the value (a factor 2 appears from
indistinguishability of the fermions)
\begin{eqnarray}
&& \hspace{-4mm}\Sigma_{\mu\nu}(p) \nn \\ && \hspace{-3mm}=
2ig^2\hspace{-1mm}\int\frac{d^2q\,dq_0}{(2\pi)^3}
q_\mu q_\nu\xi^2\left(\frac{|\q|}\Lambda\right)
G\left(\frac p2+q\right)G\left(\frac p2-q\right)
\nn \\ && \nn \\
&& 
\hspace{-3mm}
= -\delta_{\mu\nu}\frac{m^2g^2}{4\pi}\left\{\frac{\Lambda^2}{m}
+\left(p_0-\frac{p^2}{4m}\right)
\log\left(1+\frac{\Lambda^2/m}
{\frac{p^2}{4m}-p_0}\right)\right\}.
\nn \\ &&
\end{eqnarray}
All singularities in the complex $p_0$ plane lie slightly below the
real axis. The molecular propagator becomes
\begin{equation}
D(p) = \frac1{p_0-\frac{p^2}{4m}-\epsilon_0'
+c(p_0-\frac{p^2}{4m})\log\left(1+\frac{\Lambda^2/m}
{\frac{q^2}{4m}-p_0}\right)}.
\end{equation}
Here,
\begin{equation}
\epsilon_0'=\epsilon_0-\frac{mg^2\Lambda^2}{4\pi},
\end{equation}
is a renormalized detuning, while
\begin{equation}
c=\frac{m^2g^2}{4\pi}
\end{equation}
is the constant controlling the strength of the Feshbach resonance, equivalent to $c_2$ in the 3D calculations. 

Just as in 3D, the molecular propagator has a pole at $\p = 0$ and
when $p_0$ is taken to an appropriate value. We denote the real part
of the value of $p_0$ at the pole as $\omega_0$.  The pole corresponds
to the binding energy of the molecule if $\omega_0<0$ (in which case
the pole occurs at $p_0=\omega_0$) and to the resonance if
$\omega_0>0$ (in which case only the real part of $p_0$ is equal to
$\omega_0$ at the pole, while the imaginary part of $p_0$ describes
the decay rate of positive binding energy molecules in free space).
The value of $\omega_0$ is controlled by tuning $\epsilon_0$.

We can calculate $\omega_0$ from the condition that it is the real
part of $p_0$ at the pole of the molecular propagator. This gives
\begin{equation} \label{eq:41}
  \epsilon_0'= \omega_0+c\,\omega_0\log\left|1-\frac{\Lambda^2}
{m\omega_0}\right| .
\end{equation}
Then the (physical) molecular propagator is
\begin{widetext}
\begin{equation} \label{eq:42}
D(p,p_0+\omega_0) = \frac1{\left(p_0-\frac{p^2}{4m}\right)\left[1
+c\log\left(1+\frac{\Lambda^2/m}{p^2/4m-p_0-\omega_0}+i0\right)\right]
+ c\,\omega_0 \left[\log\left(1+\frac{\Lambda^2/m}{p^2/4m-p_0-\omega_0}+i0\right)
-\log\left|1-\frac{\Lambda^2}{m\omega_0}\right|\right]}.
\end{equation}
\end{widetext}
Notice that for future convenience  we shift the energy $p_0$ by $\omega_0$ so that $p_0$ in \rfs{eq:42} measures energy from the binding
energy. 

\begin{figure}[ht]
\includegraphics[height=.65 in]{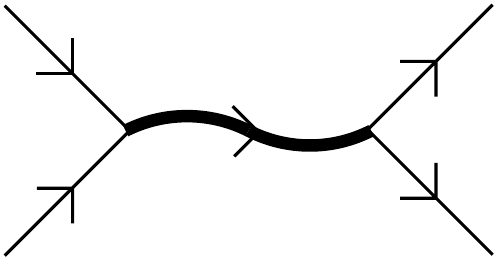}
\caption{The diagram corresponding to the scattering of two atoms.}
\label{fig:fbf}
\end{figure}

Now we can compute the scattering amplitude of two atoms in 2D, just like we did it in 3D in \rfs{eq:17}. The scattering of two atoms with
incoming momenta ${\bf k}$ and $-{\bf k}$ into momenta ${\bf k}'$ and $-{\bf k}'$ proceeds via formation of a molecule, see Fig.~\ref{fig:fbf}. Thus the scattering two-atom $T$-matrix coincides with the propagator $D_{\mu \nu}$ computed at momentum $p=0$, and at energy
$p_0+\omega_0 = k^2/m$, contracted with the incoming momentum $k_{\mu}$ and the outgoing momentum $k'_{\nu}$. 
The relationship between the scattering $T$-matrix and the scattering amplitude depends on the dimensionality of space. In 2D it is given by (see
Appendix A for a derivation)
\be \label{eq:43}
f = -\frac{m}{2 \sqrt{2 \pi k}} T.
\ee
Putting this all together gives for the partial amplitude of $p$-wave scattering, defined in \rfs{eq:b2}, 
\be \label{eq:44}  f_1(k) =  \left[ \frac {m  \sqrt{ k}}{ c \sqrt{2 \pi} } \left\{\frac{\epsilon_0'}{k^2} - \frac{1}{m} \left( c \log \frac{\Lambda^2} {k^2} +1 \right) \right\} -i \sqrt{\frac {\pi k } 2} \right]^{-1}\hspace{-2mm}.
\ee
Here $\epsilon_0'$ can be substituted in terms of $\omega_0$ using \rfs{eq:41}, and it is assumed that $\Lambda^2/k^2 \gg 1$. 
This expression conforms to the general form of the 2D scattering amplitude \rfs{eq:b4}. We emphasize that even though it was derived 
from the 2D version of the two-channel model \rfs{eq:H}, only the parameters $\epsilon_0'$ and $c$ follow from that model. Other than that, 
any $p$-wave two dimensional scattering at low energy must take this form, regardless of the model used (a similar point for the 3D scattering 
was emphasized in Ref.~\cite{Gurarie2007}). This argument will become important in section \ref{sec:IV}. 

The scattering amplitude \rfs{eq:44} has a pole at $k^2/m = \omega_0$ at negative $\omega_0$ and at real part of $k^2/m$ equal to $\omega_0$
at positive $\omega_0$, just as its 3D counterpart and as follows from the properties of the molecular propagator. 

\subsection{Stability of condensates with large molecules}
\label{sec:IIIB}

The radial part of the wave function of a 2D $p$-wave molecule with energy close to zero is given by
\be \Psi(r) \sim \frac{1}{r}.
\ee
To see this, compare with the 3D case described by Eqs.~\rf{eq:24} and \rf{eq:25} and recall that the zero energy solutions
of the Schr\"odinger equation go as $1/r^{l+1}$ in 3D and $1/r^l$ in 2D, where $l$ is the angular momentum. 
The normalization condition now follows from
\be
\int_{R_e}^{1/\kappa} r dr \left| \Psi(r) \right|^2 =  \log \left[ \frac{1}{\kappa R_e} \right],
\ee
where $\kappa=\sqrt{m \left| \omega_0 \right|}$, and where, as before, $\omega_0<0$ is the binding energy of the molecule. 
This integral is now  divergent logarithmically at both lower and upper limits, so the molecular weight is equally distributed
between $r_e$ and $1/\kappa$. So the size of the molecule is no longer $R_e$, like it was in 2D, but rather $1/\kappa$. 

To compute the rate of the atom-molecule relaxation, we go through the
same steps as we did in section \ref{sec:IIB}. However, we need to
take into account that the molecules no longer have size $R_e$, but
rather $1/\kappa$.  Let us do it, for generality reasons, in an
arbitrary number of dimensions $d$.

The decay rate is still given by 
\be
\Gamma \sim n \sigma v,
\ee where $n$ is the density, $\sigma$ is the inelastic cross section, and $v$ is the velocity,
just as in 3D, \rfs{eq:28}. However, the meaning of the inelastic scattering cross section $\sigma$ is now different. We now have two objects of
the size $1/\kappa$ colliding inelastically. The inelastic cross section is given by the product of their elastic cross section, proportional to $1/\kappa^{d-1}$, 
the time the molecules spent together, given by $1/(\kappa v)$,  the collapse rate during that time, $1/(m R_e^2)$. This should still be multiplied by the
probability that the three atoms out of four which constitute 2 molecules find themselves at distance $R_e$ from each other, so that they were all at distances of the order of the force range between them. To find this probability is not a simple problem. We have three fermions, each pair interacting strongly since they are close to a $p$-wave 
Feshbach resonance. 

Suppose their 3-body wavefunction is given by the following scaling
ansatz \be\label{eq:wavefromt} \psi(r) \sim r^{\gamma}.  \ee Here $r$
denotes a collective coordinate of the three particles, such as, for
example, a hyperspherical radius. \rfs{eq:wavefromt} serves as a
definition of a scaling exponent $\gamma$. Then the probability that
three fermions find themselves at a distance $R_e$ from each other is
given by \be P \sim \left( R_e \kappa \right)^{2d + 2 \gamma}.  \ee
The power of $2d$ comes about because of the phase volume of putting
two fermions at a distance $R_e$ from the third one, while $2\gamma$
arises from the behavior of the square of the wavefunction, $\left|
  \psi \right|^2 \sim r^{2 \gamma}$. Putting all these factors
together gives
\begin{eqnarray} \label{eq:48}
\Gamma &\sim& n \left( \frac 1 \kappa \right)^{d-1} \frac{1}{m R_e^2} \frac{1}{\kappa v} \left(R_e \kappa \right)^{2d+2\gamma} v \cr &=& \frac{1}{m \ell^d
\kappa^{d-2}} \left( R_e \kappa \right)^{2d + 2 \gamma -2}.
\end{eqnarray}
Finally, close to Feshbach resonance, the size of the molecules is close to their separation, or $1/\kappa \sim \ell$. Then the \rfs{eq:48} simplifies to give
\be \label{eq:49}
\Gamma \sim \frac{1}{m \ell^2} \left( \frac{R_e}{\ell} \right)^{2d + 2 \gamma- 2}.
\ee
This is the final answer for the decay rate of large molecules.

Let us check that this expression indeed gives the correct answer in the case of 3D $s$-wave molecules (which are large). In this case, $d=3$, and $\gamma
\approx -0.22$ \cite{Petrov2004,Petrov2005}. Then 
\be \label{eq:52}
\Gamma \sim \frac{1}{m \ell^2} \left( \frac{R_e}{\ell} \right)^{3.55},
\ee as was indeed derived in Ref.~\cite{Petrov2005}, and as was discussed in the Introduction \rfs{eq:2}. For easier comparison with Ref.~\cite{Petrov2005}, recall that it is sometimes beneficial to introduce the decay constant $K=\Gamma/n$ introduced in \rfs{eq:35}. Recall also that in the 3D $s$-wave problem, $\kappa=1/a$, where $a$ is the scattering length. This gives
\be
K \sim \frac{ R_e}{m} \left( \frac{R_e}{a} \right)^{2.55},
\ee the form discussed in Ref.~\cite{Petrov2005}. 

We also briefly examine the case of 3D $s$-wave bosons close to Feshbach resonance. Then $\gamma=-2$ (as a consequence of the
presence of Efimov states) and \rfs{eq:49} gives
\be
\Gamma \sim \frac{1}{m\ell^2}.
\ee
A decay constant for the boson problem is defined as $K = \Gamma/n^2$. Substituting the scattering length $a$ for $\ell$ we find
\be
K \sim \frac{a^4}{m},
\ee a well known form for the boson problem \cite{Fedichev1996,Nielsen1999,Greene1999,Braaten2000}.

When applied to the 2D $p$-wave problem, \rfs{eq:49} gives
\be \label{eq:53}
\Gamma \sim \frac{1}{m \ell^2} \left( \frac{R_e}{\ell} \right)^{2 + 2\gamma}.
\ee
At issue now is calculating the exponent $\gamma$. This calculation is
the subject of the next two subsections.

\subsection{The three-body problem}
\label{sec:IIIC}
We need to compute the three body wavefunction close to Feshbach resonance. This can be done in either co-ordinate or momentum space.
We are going to present the momentum space derivation, since it can be done using the standard techniques of many-body theory. 

\begin{figure}[ht]
\includegraphics[height=.65 in]{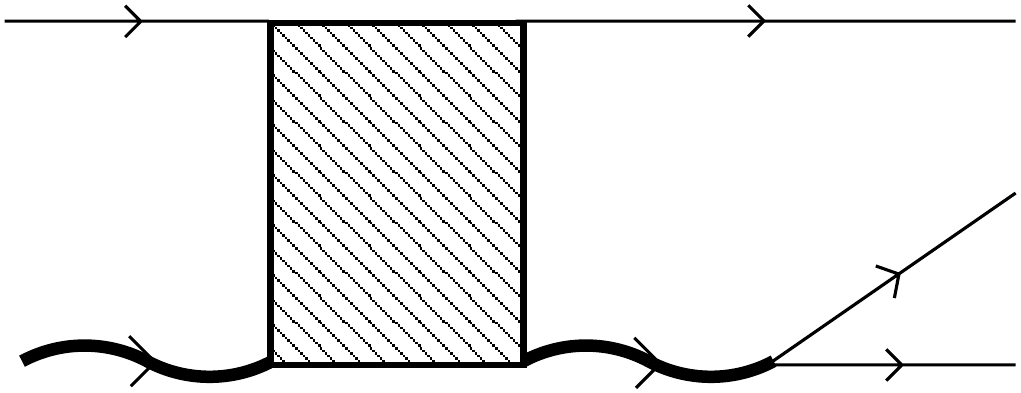}
\caption{The diagram corresponding to the wave function of free fermions. The square block represents the atom-molecule $T$-matrix,
and the three outgoing fermionic lines represent the three fermions whose wave function we would like to compute.}
\label{fig:ttowave}
\end{figure}

First of all, let us show that the scaling of the three body wave function is related to the behavior of the 3-body scattering amplitude. Suppose 
a three body scattering matrix $T$, depicted in Fig.~\ref{fig:ttowave}, is known. It is a function of the incoming momenta and energy, and the outgoing momenta and energy. To arrive 
at the wave function of three particles, we fix the incoming momenta and energy to be on shell. Then we multiply the $T$ matrix by the four outgoing
Green's functions, one bosonic and three fermionic. Finally we Fourier transform with respect to the fermionic outgoing momenta, respecting the momentum-energy conservation. For the illustration of this procedure, see \rfs{eq:a1} which represents this in case of just one particle scattering off a potential. 

Suppose the $T$-matrix scales as $p^{\gamma'}$, where $p$ represents the overall momentum scale. Then it is easy to figure out the scaling of the wave function. There are three outgoing energies and momenta, corresponding to the three outgoing fermionic lines. However, the energy-momentum conservation restricts the number of linear independent energies and momenta to two. Hence there are two integrals over energy and momentum, contributing the power $2 (d+2)$ (energy is counted as momentum squared). There are four outgoing propagators, one bosonic \rfs{eq:42} and three fermionic \rfs{eq:34}, contributing the power $-8$ (the bosonic propagator includes logarithms, but these are irrelevant for the purpose of power counting). Finally there is a vertex where a bosonic line splits into two fermions, contributing a power $1$ due to the $p$-wave momentum dependent factor. Putting it all together, and remembering that the coordinate scaling is opposite in sign to the momentum scaling, we find that the scaling of the wave function is given by
\be \gamma = - \left( \gamma' + 2(d+2) - 8+1 \right),
\ee
or in two dimensions,
\be \label{eq:55} \gamma = -\gamma'-1.
\ee

To proceed, we need to know the scaling of the $T$-matrix, representing
 the scattering of a fermion and a spin 1
molecule.

\subsection{Solution of the three body problem}
\label{sec:IIID}

In the scattering problem, let the incoming molecule have spin $\mu$
and the outgoing spin $\nu$. The $T$-matrix will then in general be a
tensor $T_{\mu\nu}$. In the center of mass frame, with the incoming
molecule having momentum $\k$ and the outgoing $\p$, the tensor
$T_{\mu\nu}$ consists of five terms proportional to $\delta_{\mu\nu},$
$p_\mu p_\nu,$ $p_\mu k_\nu,$ $k_\mu p_\nu,$ and $k_\mu
k_\nu$. However, we are interested in the wave function at short range $r \ll 1/\kappa$,
which corresponds to large $p$, or $p \gg k \sim \kappa$. Then it is sufficient to set ${\bf k}=0$.  Thus define
\cite{Levinsen2007a}
\begin{eqnarray}
T_{\mu\nu}(\p,p_0) & \equiv & T_1(p,p_0)\delta_{\mu\nu}+T_2(p,p_0)p_\mu p_\nu/p^2
\nn \\ & \equiv & \sum_{i=1,2}T_i(p,p_0)u^i_{\mu\nu}(\vec p).
\label{eq:basistensor}
\end{eqnarray}

\begin{figure}[ht]
\includegraphics[height=.34 in]{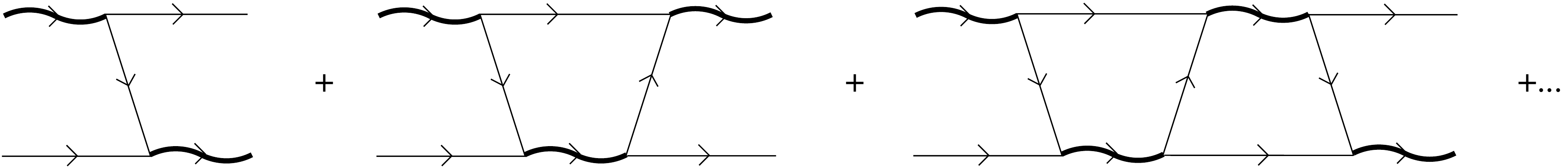}
\caption{The series of diagrams which sum to give the scattering
  $T$-matrix of a spin 1 molecule and a fermionic atom.}
\label{fig:firstfb}
\end{figure}

The scattering $T$-matrix is the sum of the series of diagrams
indicated in Fig. \ref{fig:firstfb}. These diagrams appear the same as
those studied in the $s$-wave three-body problem
\cite{Skorniakov1956,Bedaque1999}, however the Feynman rules are of
course different in the present problem. For weak resonances, the
diagrams form a perturbative series in which only the first few
diagrams may be kept. However, for the strong resonances studied in
the present paper the diagrams are all of the same order and the sum
of the series of diagrams is needed. The summation may be attained by
constructing a Lippmann-Schwinger type integral equation for the
scattering amplitude. This summation was performed in a similar manner
in the three-dimensional problem studied in Ref. \cite{Levinsen2007}
(see also Ref. \cite{Lasinio2008}). The kinematics are chosen as
follows; the incoming molecule is on-shell with three-momentum $({\bf
  0},\omega_0)$ while the incoming atom has three-momentum $({\bf
  0},0)$. The outgoing molecule has $(\p,p_0+\omega_0)$ and the
outgoing fermion $(-\p,-p_0)$. The scattering matrix does not include
external lines. Then the integral equation takes the form
\begin{eqnarray}
&& \hspace{-4mm}T_{\mu\nu}(\vec p,p_0) = -2Z\,g(p)g(p/2)
G(p,p_0+\omega_0)p_\mu p_\nu 
\nn \\ &&
\hspace{-3mm}-4i\hspace{-1mm}\int
\frac{d^2q\,dq_0}{(2\pi)^3}T_{\mu\alpha}(\q,q_0)
G(q,-q_0)G(\p+\q,p_0+q_0+\omega_0)
\nn \\ && \hspace{-1mm}
\times D(q,q_0+\omega_0)(p+\frac q2)_\alpha(q+\frac p2)_\nu
 g(|\p+\frac \q2|)g(|\q+\frac \p2|).
\nn \\
\label{eq:2dinteq}
\end{eqnarray}
Repeated indices are summed over. $Z$ is the residue of the molecular
propagator at the pole and is needed for correct normalization of the
scattering matrix. It is a function of $\omega_0$ and $c$ whose
precise value will not be needed in the following.

The integral over $q_0$ in Eq. (\ref{eq:2dinteq}) may be performed by
closing the contour in the upper half plane, setting $q_0\to
-q^2/2m$. In order to solve the integral equation it is then
convenient to let $p_0\to -p^2/2m$. This ensures that the frequency
dependence of $T_{\mu\nu}$ is the same on both sides of the integral
equation. The integral equation is then solved for $T_i(p)\equiv
T_i(p,-p^2/2m)$. Subsequently, this solution can then be used to find
$T_i(p,p_0)$ at any $p_0\leq0$.

To project onto the functions $T_1$ and $T_2$ defined in
Eq. (\ref{eq:basistensor}) multiply the integral equation
(\ref{eq:2dinteq}) by $u^k_{\mu\nu}(\p)$. The left hand side will
then contain the matrix
\begin{equation}
U_{ki}=u^k_{\mu\nu}(\p)u^i_{\mu\nu}(\p)=\left(\begin{array}{cc}2&1\\1&1
\end{array}\right)_{ki}.
\end{equation}
This matrix is invertible and it is thus possible to find a set of
coupled integral equations for the functions $T_1(p)$ and
$T_2(p)$. Upon multiplying Eq. (\ref{eq:2dinteq}) by
$U^{-1}_{jk}u^k_{\mu\nu}(\p)$ it becomes
\begin{widetext}
\begin{equation}
T_j(p) = -2Z\,g(|\p|)g(|\p|/2)
G(p,-p^2/2m+\omega_0)p^2\delta_{2j}
-\frac{mg^2}{\pi^2}\int q\,dq\,D(q,-q^2/2m+\omega_0)b_{ji}(p,q)T_i(q,-q^2/2m).
\label{eq:t1t2}
\end{equation}
The dimensionless matrix $b$ is given by 
\begin{equation}
b_{ji}(p,q)\equiv \frac1{m}U^{-1}_{jk}\int_0^{2\pi}d\theta\,
\frac{\xi(|\p+\q/2|)\xi(|\q+\p/2|)}{\omega_0-p^2/m-q^2/2-\p\cdot\q/m}
u^k_{\mu\nu}(\p)u^i_{\mu\alpha}(\q)(p+q/2)_\alpha
(q+p/2)_\nu,
\end{equation}
with $\theta$ the angle between $\p$ and $\q$.


To further simplify, assume that both $p$ and $q$ are cut off at
$\Lambda$ using the usual step function. Then the matrix $b$ may be
calculated with the result
\begin{eqnarray}
b_{11} & = & \frac\pi{p^2}\left[\sqrt{(\omega-p^2-q^2)^2-p^2q^2}
+\omega-p^2-q^2\right], \\
b_{12} & = & -\pi+\pi\frac{(2\omega-2p^2-q^2)(\omega-p^2-q^2)}{p^2q^2}
\nn\\&&+\frac\pi{\sqrt{(\omega-p^2-q^2)^2-p^2q^2}}\left[
\frac{(2\omega-2p^2-q^2)(\omega-p^2-q^2)^2}{p^2q^2}-2\omega+2p^2+q^2\right], \\
b_{21} & = & -\frac{5\pi}{2}+\frac\pi2\frac{3p^2+3q^2-5\omega}{\sqrt{(\omega-p^2-q^2)^2
-p^2q^2}} -\frac{2\pi}{p^2}\left[\sqrt{(\omega-p^2-q^2)^2-p^2q^2}+\omega
-p^2-q^2\right], \\
b_{22} & = & -\frac\pi2-\pi\frac{(4\omega-3p^2-2q^2)(\omega-p^2-q^2)}{p^2q^2}
\nn\\&&+\frac{\pi/2}{\sqrt{(\omega-p^2-q^2)^2-p^2q^2}}\left[-2
\frac{(4\omega-3p^2-2q^2)(\omega-p^2-q^2)^2}{p^2q^2}+3\omega-3p^2-q^2\right].
\label{eq:matrixb}
\end{eqnarray}
Here a dimensionless detuning is defined as $\omega\equiv
\omega_0\frac m{\Lambda^2}$.
\end{widetext}

According to Eq. (\ref{eq:wavefromt}), in order to determine the
behavior of the three-body wavefunction, the functions $T_1(p)$ and
$T_2(p)$ are needed for momenta $\sqrt{-m\omega_0}\ll p\ll\Lambda$. We
now proceed to solve Eq. (\ref{eq:t1t2}) analytically in this range of
momenta. In the following, the limit $c\to\infty$ will be taken to
ensure that the Feshbach resonances studied are strong. For momenta in
the range of interest the integral equation reduces to
\begin{equation}
T_j(p) = 2Zg^2m\delta_{2,j}+\frac8{3\pi}\int_{\kappa}^\Lambda
\frac{dq}q\frac{b_{ji}(p/q)}{\log\left(\frac\Lambda q\right)}T_i(q).
\label{eq:tapprox}
\end{equation}
The matrix $b$ takes the limiting forms
\begin{eqnarray}
&&\hspace{-25mm}
\kappa \ll p\ll q \ll\Lambda: \nn \\
&&\hspace{-15mm}
b = \left(\begin{array}{cc} -\frac\pi2+\frac{3\pi}8\frac{p^2}{q^2}
& -\frac\pi2+\frac{5\pi}8\frac{p^2}{q^2} \\
-\frac{9\pi}{16}\frac{p^4}{q^4} & -\frac{3\pi}{16}\frac{p^4}{q^4}
\end{array}\right) \\
&&\hspace{-25mm}
\kappa\ll q\ll p \ll\Lambda: \nn \\
&&\hspace{-15mm}
b = \left(\begin{array}{cc} -\frac\pi2\frac{q^2}{p^2}
& -\frac\pi4\frac{q^2}{p^2} \\
-\pi+\frac{7\pi}4\frac{q^2}{p^2} & -\frac\pi2+\frac{7\pi}8\frac{q^2}{p^2}
\end{array}\right)
\end{eqnarray}
Eq. (\ref{eq:tapprox}) suggests a solution of the form
\begin{equation}
T_1(p) = C_1\log^\alpha\left(\frac\Lambda p\right),
\hspace{5mm}
T_2(p) = C_2\log^\alpha\left(\frac\Lambda p\right).
\end{equation}
Substituting into Eq. (\ref{eq:tapprox}) and keeping leading terms
results in the set of linear equations for the coefficients $C_1$ and
$C_2$
\begin{eqnarray}
&&\hspace{-5mm}C_1\log^\alpha\left(\frac\Lambda p\right) =
-\frac4{3\alpha}\left\{C_1\log^\alpha\left(\frac\Lambda p\right)
+C_2\log^\alpha\left(\frac\Lambda p\right)\right\} \nn \\
&& \hspace{-5mm}
C_2\log^\alpha\left(\frac\Lambda p\right)
= -\frac4{3\alpha}\left\{2C_1\left[
\log^\alpha\left(\frac\Lambda\kappa\right)
-\log^\alpha\left(\frac\Lambda p\right)\right] \right. \nn \\
&&\left.\hspace{14mm}
+
C_2\left[\log^\alpha\left(\frac\Lambda\kappa\right)-
\log^\alpha\left(\frac\Lambda p\right)\right]\right\}
+2Zg^2m. \nn\\ &&
\end{eqnarray}
This set of equations has solutions only if
$\alpha=\pm\frac43i$. Matching coefficients finally results in the
solutions
\begin{eqnarray}
T_1(p) & = & \rho\cos\left[\frac43\log\log\left(\frac\Lambda p\right)
+\phi+\frac{3\pi}4\right], \nn \\
T_2(p) & = & \sqrt2\rho\cos\left[\frac43\log\log\left(\frac\Lambda p\right)
+\phi\right],
\label{eq:t1t2an}
\end{eqnarray}
with the amplitude satisfying
\begin{equation}
\rho = \frac{\sqrt2Zg^2m}
{\cos\left[\frac43\log\log\left(\Lambda/\kappa\right)\right]}.
\end{equation}
This solution contains a free parameter, as it does not allow the
determination of both $\rho$ and $\phi$ independently.

In solving Eq. (\ref{eq:t1t2}) numerically, it is found that the
overall amplitude of the solutions $T_1(p)$ and $T_2(p)$ converges
very slowly. We attribute this slow convergence to the logarithmic
behavior of the solutions; in the numerical study it is important to
keep $\log(\Lambda/\kappa)\gg1$ while simultaneously the configuration
space must contain a large number of momenta $p$ for which
$\log(\kappa/p)\gg1$. Having this in mind, we therefore write the
solutions as
\begin{equation}
T_i(p) = \rho_i\cos\left[\frac43\log\log\left(\frac\Lambda p\right)
+\phi_i\right],\hspace{1cm} i=1,2
\label{eq:t1t2an2}
\end{equation}
and determine the ratio $\rho_2/\rho_1$ rather than the separate
values of these amplitudes.

\begin{figure}[ht]
\includegraphics[height=3 in]{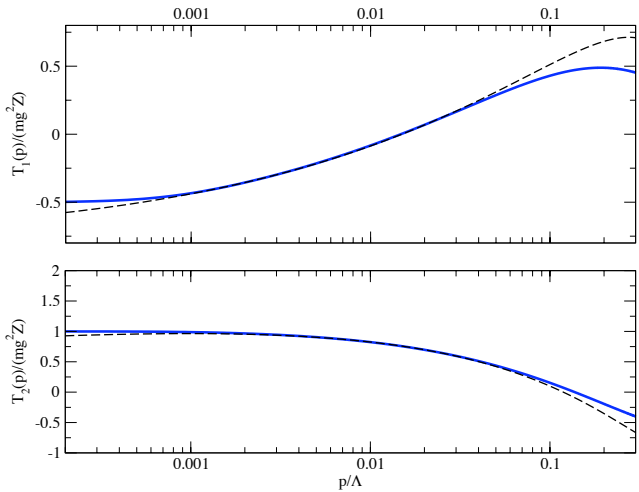}
  \caption{$T_1(p)$ and $T_2(p)$ obtained by solving
    Eq. (\ref{eq:t1t2}) for $\omega_0=-10^{-6}\frac{\Lambda^2}m$ and
    $c_2=10^6$ (blue, thick). The solution uses Gaussian-Legendre
    quadrature for the numerical integration \cite{numrecipes} with
    2000 grid points. Also shown are the analytical solutions
    (\ref{eq:t1t2an2}) with parameters chosen as in
    Eq. (\ref{eq:solnpar}) (black, dashed).}
\label{fig:t1t2}
\end{figure}

The solutions $T_1(p)$ and $T_2(p)$ of the coupled integral equations
(\ref{eq:t1t2}) are shown in Fig. \ref{fig:t1t2} for a large value of
$c_2$. Also shown are the analytical solutions (\ref{eq:t1t2an2}) with
the parameters
\begin{eqnarray}
\phi_1 = 2.80 && \phi_2 = .56 \nn \\
\rho_1 = .71 && \rho_2 = .96\,. \label{eq:solnpar}
\end{eqnarray}
We observe that $\phi_1-\phi_2\approx 3\pi/4$ and
$\rho_2/\rho_1\approx\sqrt2$, both with a $5\%$ error. Thus we
conclude that the analytical expressions given in
Eq. (\ref{eq:t1t2an}) are correct.

\subsection{The lifetime and the interaction energy of the 2D $p$-wave condensates}
\label{sec:IIIE}
We are now in a position to finish the calculation of the lifetime of the 2D $p$-wave condensate. 
The scattering amplitude is given by Eq. (\ref{eq:basistensor}), where in turn $T_1$ and $T_2$ are given by
\rfs{eq:t1t2an}. These expressions do not scale at all, corresponding to $\gamma'=0$. Thus, according to
\rfs{eq:55}, \be \gamma=-\gamma'-1=-1. \ee
Substituting this into \rfs{eq:53} we find
\be\label{eq:76}
\Gamma \sim \frac{1}{m \ell^2},
\ee
which gives the decay rate of the 2D $p$-wave superfluid as discussed in the Introduction, in \rfs{eq:3}. 

This should be compared with the interaction rate for the 2D $p$-wave atoms at the Fermi energy of the gas. The find this energy 
first the BCS regime by examining the scattering 
amplitude of two atoms at positive $\omega_0$, as found earlier, \rfs{eq:44}.  The imaginary part of the pole of this scattering amplitude, computed at $\omega_0=\epsilon_F$, gives us the  needed interaction rate, quite analogously with \rfs{eq:30} which we employed in 3D.  The answer crucially
depends on whether the parameter
\be \label{eq:77} c \log \left[\frac { \ell}{R_e} \right] \ee  is large or small (as before, $\ell$ is the interparticle separation and $R_e=1/\Lambda$ is the interaction length scale). 
As in the 3D case, we can term the case when this parameter is large as strong resonance, although unlike in 3D, any resonance when the gas is sufficiently dilute becomes strong. If the resonance is weak, then $c \ll 1$, since the logarithm in \rfs{eq:77} is always large. 

Assuming that the resonance is strong, we find the pole of \rfs{eq:44} to be at
\be
\frac{k^2}{m} \approx \omega_0 - i \frac{\pi \omega_0}{\log \left[ \frac{\Lambda^2}{m \omega_0} \right]}.
\ee
In the opposite case of weak resonance, we find
\be
\frac{k^2}{m} \approx \omega_0 - i \pi c \omega_0.
\ee
Substituting $\omega_0=\epsilon_F$, we estimate from here the atomic interaction energy as
\be \label{eq:80}
E_{2D} \sim \frac{1}{m \ell^2} \frac{1}{\log \left[ \frac{\ell}{R_e} \right]}
\ee
in the case when the resonance is strong and 
\be \label{eq:81}
E_{2D} \sim  \frac{c}{m \ell^2} 
\ee
when the resonance is weak. We expect that in a typical experiment the gas will be dilute so the resonance will be strong, and that's why we quoted
\rfs{eq:80} in the Introduction, \rfs{eq:6}. In either of these two cases, we see that the interaction energy is much smaller than the decay rate, given by 
\rfs{eq:76}, so we expect purely 2D $p$-wave gases to be unstable. 

Likewise, in the BEC regime, we need to estimate the elastic scattering rate of two molecules. The molecules (unlike atoms) scatter
in the $s$-wave channel. The scattering  of two molecules in 2D must proceed
according to the standard rules of quantum mechanics, which predicts that the $s$-wave scattering cross section of particles at sufficiently low momentum $k$ goes
as \cite{LL}
\be 
\sigma_{el} = \frac{\pi^2}{k} \frac{1}{\log^2 \left( \frac{ \alpha R_e}{k} \right) + \frac{\pi^2}4},
\ee
where the unknown constant $\alpha$ depends on the details of the interactions. This gives for the elastic collision rate (taking $k$ of the
order of $1/\ell$)
\be \label{eq:2DBEC}
E_{2D, BEC} = \sigma_{el} n v \sim \frac{\pi^2}{m \ell^2} \frac{1}{\log^2 \left( \frac{ \alpha R_e}{k} \right) + \frac{\pi^2}4},
\ee
This energy is somewhat smaller than the BCS estimate \rfs{eq:80}, but since the logarithmic factor is not very large, we can think of it
as being of the same order as \rfs{eq:80}. Thus the BEC 2D superfluid is as unstable as its BCS counterpart. 

\section{Quasi-2D superfluid}
\label{sec:IV}
Now let us consider the last remaining question, the stability of a 2D $p$-wave superfluid confined to a pancake of width $d$. Such a superfluid is still described by \rfs{eq:H}, but with a presence of an extra confining potential in the third direction, which we denote $z$
\be
V(z) = \oh m \omega^2 z^2,
\ee
where $\omega$ is the confining frequency. The width of the pancake $d$ is related to the oscillator frequency via
\be
d\sim \frac 1 {\sqrt{m \omega}}.
\ee
The quasi-2D regime is \be R_e \ll d \ll \ell.\ee
$R_e \ll d$ because otherwise the physics of the Feshbach resonance is modified by the confinement (also, $R_e$ is very short so that the confinement to the scale below $R_e$ is not currently technologically possible), while $d \ll \ell$ in order for the gas to be truly confined to 2D.  

The scattering of identical fermions close to a $p$-wave Feshbach resonance follows from this Hamiltonian. Calculating  it in the confined geometry is an involved problem, first
solved for the case of $s$-wave resonance confined to 2D in Ref.~\cite{Petrov2000,Petrov2001} (see also the first calculation of this type, done for the $s$-wave gas confined to 1D, 
in Ref.~\cite{Olshanii1998}). For the case of $p$-wave resonances, this problem was studied in  Ref.~\cite{Pricoupenko2008} in both 1D and 2D.
Yet, for our purposes, we do not need to know the answer. It is enough to know that the scattering is still described by \rfs{eq:44}, albeit with coefficients $c$ 
and $\epsilon_0'$ no longer related to the parameters of the Hamiltonian \rfs{eq:H} the way they were before, but rather being some more complicated functions, which also depend on $d$ among other parameters. This is because any low energy $p$-wave scattering in two dimensions must
be described by the expression \rfs{eq:b4}, with the function $g_1(k)$ having the low $k$ expansion as in \rfs{eq:44}. 

Thus the interaction energy of the two atoms confined to this geometry, derived in the previous section by using \rfs{eq:44} only, is still given by \rfs{eq:80} or \rfs{eq:81}. The only difference is that $R_e$ in these relations should be traded for $d$, as $d$ is now the smallest lengthscale at which the 2D physics
is still at work. 

Yet the interaction rate of the molecules in the BEC regime will actually be given by the 3D formula \rfs{eq:3DBEC}. This is related to the fact that in
quasi-2D geometry, the coefficient $\alpha$ in \rfs{eq:2DBEC} is typically anomalously large, and leads to \cite{Petrov2001}
\be
E_{quasi-2D, BEC} \sim \frac{1}{m \ell^2}\frac{R_e^2}{d^2}.
\ee
This energy scale is very small, thus there is no hope to observe the BEC of $p$-wave molecules, even in the quasi-2D geometry. So we concentrate
on the case of the BCS phase. 


Now the decay rate in quasi-2D can be deduced in the following way. Atoms decay when three of them approach each other at distance $R_e$. This distance is much shorter than $d$, so that the 3D decay physics must take over. So we should not use \rfs{eq:76} to compute the decay rate. Rather, we need to revert back to the appropriate expression in 3D, given by \rfs{eq:29}, with one modification. In \rfs{eq:29}, $\ell$ denotes average distance between the particles in three dimensions, while \rfs{eq:80} and \rfs{eq:81} are written in terms of average distance between particles in two dimensions. These are related by the obvious
\be
\ell_{3D}^3 = \ell_{2D}^2 d.
\ee
This leads to the decay rate
\be
\Gamma_{quasi-2D} \sim \frac{1}{m \ell^2} \frac{R_e}{d},
\ee
where $\ell$ is now the two dimensional distance. This concludes the derivation of \rfs{eq:4}. 

We therefore see that the necessary condition for the existence of the stable superfluid in the quasi-2D geometry is given by
$E_{2D} \gg \Gamma_{quasi-2D}
$ or
\be \label{eq:87}
\max \left( \log \left[ \frac{\ell}{d} \right], \frac 1 c \right) \ll \frac{d}{R_e}.
\ee
Since logarithms, even of large arguments, are typically not very large (we assume $c$ is a constant generally of the order of $1$, although its value
is controlled purely by a particular Feshbach resonance and it can be calculated from its physics on a case by case basis), it is possible that this condition will be satisfied in experiments. 

Indeed, a typical value for $d$ would be, perhaps, half a wavelength of the light used to create a confining potential. This gives $d \sim 250$ nm, or $5\cdot 10^3$ au. With $\ell \sim 10^4$ au, the ratio $\ell/d \sim 2$, and its logarithm is basically 1.  At the same time $d/R_e$ can be kept just marginally smaller than $200$. So \rfs{eq:87} is satisfied with a large margin. 

Yet we must remember than \rfs{eq:87} is but a necessary condition for the decay being slow enough to allow for equilibration of the superfluid. In practice, the equilibration involves many collisions between atoms and may go much slower than $E_{2D}$. In truth we have only one example of the decay slow enough that that we know from experiment that there is enough time for equilibration and meaningful experiments in the superfluid phase, that of the
3D $s$-wave case. In that case, the ratio of the Fermi energy (as a crude estimate of the characteristic energy of the condensate) to the decay rate is an impressive $\left(\ell/R_e \right)^{3.55} \sim 10^8$. The $p$-wave superfluids confined to 2D are far from being that stable. 

\newpage

\section{Conclusions}
This concludes our studies of the stability of the fermionic paired superfluids close to $p$-wave Feshbach resonance. From the analysis of this paper, it is clear that  the 3D $p$-wave gases are inherently unstable. The situation improves if they are confined to 2D. Yet it is nowhere near the
case of 3D $s$-wave superfluids, which are stable for all practical purposes due to a very high scaling power in their decay rate \rfs{eq:52}.
A promising route to increase stability further in a quasi-2D setting seems to be a creative application of the optical lattice, similar to what was recently done in Ref.~\cite{Rempe2008} for a different problem. This should be a subject of further work.

\acknowledgements

We acknowledge support by NSF via grant DMR-0449521 (VG and JL), by EPSRC GR/S61263/01 (NC), and by the IFRAF Institute and ANR Grant No. 06-NANO-014 (JL). We thank D. Petrov  for useful discussions and C. H. Greene for helpful remarks.

\appendix
\section{Relationship between the $T$-matrix and the scattering amplitude in 2D}

Consider a particle of mass $m$ scattering in a potential in 2D.  Its wave function is given by
\be \label{eq:a1} \psi({\bf r}) = e^{i {\bf  k} {\bf r}} + \int \frac{d^2 p}{(2 \pi)^2}
~\frac 1 {E-\frac{p^2}{2m}+i0} T( {\bf k},{\bf p}) ~e^{i {\bf p} {\bf
r}}, \ee
where $T( {\bf k},{\bf p})$ is the scattering $T$-matrix computed between momenta ${\bf k}$ and ${\bf p}$ at energy $E=k^2/(2m)$.
We compare this expression with the definition of the scattering amplitude in 2D, given by
the large $r$ expression of the wave function \cite{LL}
\be \label{eq:a2}
\psi ({\bf r}) = e^{i {\bf k} {\bf r}}+ f(k, \varphi) \frac{e^{i k r}}{\sqrt{-i r}}, \ \sqrt{-i} = \exp \left(- i \pi/4 \right),
\ee
where $\varphi$ is the angle between the incoming momentum ${\bf k}$ and the position vector ${\bf r}$.
Doing the angular part of the
integral in \rfs{eq:a1} at large $r$ by the steepest descend method we find
\begin{widetext}
\be
\psi({\bf r}) = e^{i {\bf k} {\bf r}} +  \int_{0}^{\infty} \frac{p~ dp}{4 \pi^2}
\left( T( {\bf  k},{\bf  p}_r) \sqrt{\frac{2\pi}{i p r}} e^{i p r}  +T(
{\bf k},-{\bf  p}_r)\sqrt{\frac{2\pi}{-i p r}} e^{-i p r} \right) \frac 1 {\frac {k^2}{2m}
- \frac{p^2}{2m} + i 0},
\ee
\end{widetext}
where ${\bf  p}_r$ denotes a vector whose length is $p$, but which is directed along ${\bf r}$. 
Change variables in the second integral  to get
\be \int_{-\infty}^\infty \frac{p~dp}{4\pi^2} ~ T( {\bf k},{\bf  p}_r) \sqrt{\frac{2\pi}{i p r}} e^{i p r} \frac 1 {\frac {k^2}{2m}
- \frac{p^2}{2m} + i 0},\ee
where the contour of integration goes above the $p=0$ singularity. Doing the integral by residues gives
\be \psi({\bf r}) =  e^{i {\bf k} {\bf r}} - \frac{m i}{\sqrt{2 \pi i k r}} T( {\bf  k},{\bf  k}_r)e^{i k r}. \ee
Comparing with \rfs{eq:a2} gives
\be \label{eq:a6} f = -\frac{m}{\sqrt{2\pi k}}T({\bf k}, {\bf k}').\ee
When comparing this expression with the one used in the text, \rfs{eq:43}, one needs to remember that $m$ in \rfs{eq:a6} is the
reduced mass of two fermions and should be replaced according to $m \to m/2$.

\section{Constraints placed on the scattering amplitude in 2D by unitarity}
The scattering amplitude in 2D is constrained by unitary, just like its 3D counterpart. Here we reproduce the
appropriate derivation from Ref.~\cite{LL}. 

Consider the scattering of a particle of mass $m$ with momentum ${\bf k}$ into the momentum ${\bf k}'$ such that $k=k'$, but the angle between these two vectors is $\varphi$. The scattering amplitude can be expressed in terms of the 
phase shifts $\delta_l$ according to \cite{LL}
\be f(k,\varphi) = \frac{1}{i \sqrt{2 \pi k}} \sum_{l=-\infty}^{l=\infty} \left(e^{2 i \delta_l}- 1 \right) e^{i l \varphi}. \ee
Introduce the partial scattering amplitudes
\be \label{eq:b2} f_l(k) = \frac{1}{i \sqrt{2 \pi k}}\left(e^{2 i \delta_l}- 1 \right). \ee
Since  \be \left| e^{2 i \delta_l} \right|^2 =1, \ee a constraint on the form $f_l(k)$ can take follows,
\be \label{eq:b4} f_l(k) = \frac 1 {g_l(k) - i \sqrt{\frac{\pi k}{2}}}.
\ee
Here $g_l(k)$ are {\sl real} functions of $k$. 

Thus only the functions $g_l(k)$ remain undetermined in an arbitrary scattering process. Yet their
low $k$ expansions take a universal form, up to the coefficients of the expansion \cite{LL}. It is these coefficients which need
to be calculated on a case by case basis (see \rfs{eq:44} for the low $k$ expansion of $g_1(k)=g_{-1}(k)$).

\bibliography{paper}

\begin{thebibliography}{48}
\bibitem[{Com()}]{CommentDim}
\bibinfo{note}{The $p$-wave two channel model has an additional factor of
  momentum in its interactions compared to its $s$-wave counterpart. This
  changes the power counting effectively replacing $d$ with $d-2$, so that
  whereas the $s$-wave two channel model has upper critical dimension $d=4$,
  the $p$-wave two channel model's upper critical dimension is $d=2$.}
\end{thebibliography}

\end{document}